\documentclass[aps,pra,onecolumn,superscriptaddress,reprint,longbibliography]{revtex4-1}

\usepackage{amsmath,hyperref,graphicx,color}

\hypersetup{
    colorlinks=true,       
    linkcolor=blue,          
    citecolor=blue,        
    filecolor=blue,      
    urlcolor=blue           
}

\newsavebox{\graphicsbox}
\graphicspath{{./}{./figs/v5/}}

\usepackage{txfonts}

\def\um{\mbox{ $\mu$m}}
\def\nm{\mbox{ nm}}

\begin{document}
  \title{A rapidly expanding Bose-Einstein condensate: an expanding universe in the lab} 
  \author{S. Eckel}
  \author{A. Kumar}
  \affiliation{Joint Quantum Institute, National Institute of Standards and Technology and University of Maryland, Gaithersburg, Maryland 20899, USA}
  \author{T. Jacobson}
  \affiliation{Department of Physics, University of Maryland, College Park, Maryland 20742, USA}
  \author{I.B. Spielman}
  \author{G.K. Campbell}
  \affiliation{Joint Quantum Institute, National Institute of Standards and Technology and University of Maryland, Gaithersburg, Maryland 20899, USA}

\begin{abstract}
We study the dynamics of a supersonically expanding ring-shaped Bose-Einstein condensate both experimentally and theoretically.  The expansion redshifts long-wavelength excitations, as in an expanding universe.  After expansion, energy in the radial mode leads to the production of bulk topological excitations -- solitons and vortices -- driving the production of a large number of azimuthal phonons and, at late times, causing stochastic persistent currents.  These complex nonlinear dynamics, fueled by the energy stored coherently in one mode, are reminiscent of a type of ``preheating'' that may have taken place at the end of inflation.
\end{abstract}

\maketitle

Cosmological expansion is central to our understanding of the universe.  Here, we experimentally create a system where fields expand in a similar way as in the universe: an expanding, ring-shaped atomic Bose-Einstein condensate (BEC).   Our laboratory test bed demonstrates three effects associated with the expanding universe.  First, we conclusively demonstrate a redshifting of phonons analogous to the redshifting of photons, which provided the first evidence for an expanding universe~\cite{peebles1993principles}.  Second, we observe hints of ``Hubble friction'' that damps the redshifted fields~\cite{Baunmann2012}.  Third, we observe a process in which energy is rapidly transferred from a homogeneous radial mode into azimuthal modes by a nonlinear, turbulent cascade, reminiscent of that seen in some models of preheating at the end of cosmological inflation~\cite{Kofman1994,Felder2001,Kofman2001,Copeland2002,Gleiser2010,Amin2012,Amin2014}. Experiments such as these can  thus emulate both linear and nonlinear field theoretic aspects of cosmology.

A zero-temperature BEC is a vacuum for phonons~\cite{pitaevskii2003bose}, just as an empty universe is a vacuum for quantum fields, like photons.   In this analogy, the speed of light is replaced by the speed of sound, $c$, in the BEC.   Evoking general relativity, the equation for long-wavelength phonons (in the hydrodynamic limit~\cite{Note1}) takes on a covariant form with a curved spacetime metric~\cite{Unruh1981,Fischer2002,Barcelo2005}.  Previous studies with ultra-cold atoms illuminated different aspects of this phonon metric.  For example, an interface between regions of sub-sonic and super-sonic fluid flow forms a ``sonic event horizon'' that exhibits effects such as Hawking radiation~\cite{Unruh1981,Garay2000,Garay2001,Barcelo2005,Unruh2007, Jain2007,Macher2009,Lahav2010,Steinhauer2014,Steinhauer2016}.  By changing the interaction strength or density, one can simulate cosmological phenomena such as pair production~\cite{Fedichev2003,Barcelo2003,Fedichev2004,Fischer2004,Calzetta2005,Jain2007a,Prain2010}, Sakharov oscillations~\cite{Sakharov1966}, or the dynamical Casimir effect~\cite{Moore1970}, the latter two having been recently observed experimentally~\cite{Hung2013,Jaskula2012}.  Beyond cold atoms, experimental studies have realized analog event horizons in other settings, for example in optical systems~\cite{Belgiorno2010,Rubino2012,Vocke2017} and in classical fluids~\cite{Weinfurtner2011,Euve2016,Torres2017}.  (For a recent review, see Ref.~\cite{Faccio2013}.)

The expansion of our BEC-universe is forced by dynamically increasing the radius of our nearly-flat bottomed ring-shaped potential~\cite{Note2}, as opposed to being governed by an analog of the Einstein equation (see Appendix~\ref{sec:exp_details}).  Figure~\ref{fig:intro} shows our BEC during a $t_\text{exp} = 23.1$~ms inflation.  The radial velocity of the trapping potential (defined as the rate of change of the mean radius, $R$) is directly controlled, and can be made comparable to the speed of sound.  For the expansion shown in Fig.~\ref{fig:intro}, the maximum velocity is $v_p = dR/dt \approx 1.3 c$, implying that points separated by an angle $\gtrsim \pi/4$ recede faster than $c$.  The condensates used in this work are well-described by mean field theory; thus, we compare our measurements to numerical simulations using the stochastic-projected Gross-Pitaevskii equation (SPGPE, see Appendix~\ref{app:spgpe}), which accurately captures BEC dynamics with thermal fluctuations~\cite{Rooney2012,Bradley2015}. Images from this simulation are in excellent agreement with the corresponding experimental images.

\onecolumngrid

\begin{figure*}[h]
	\includegraphics{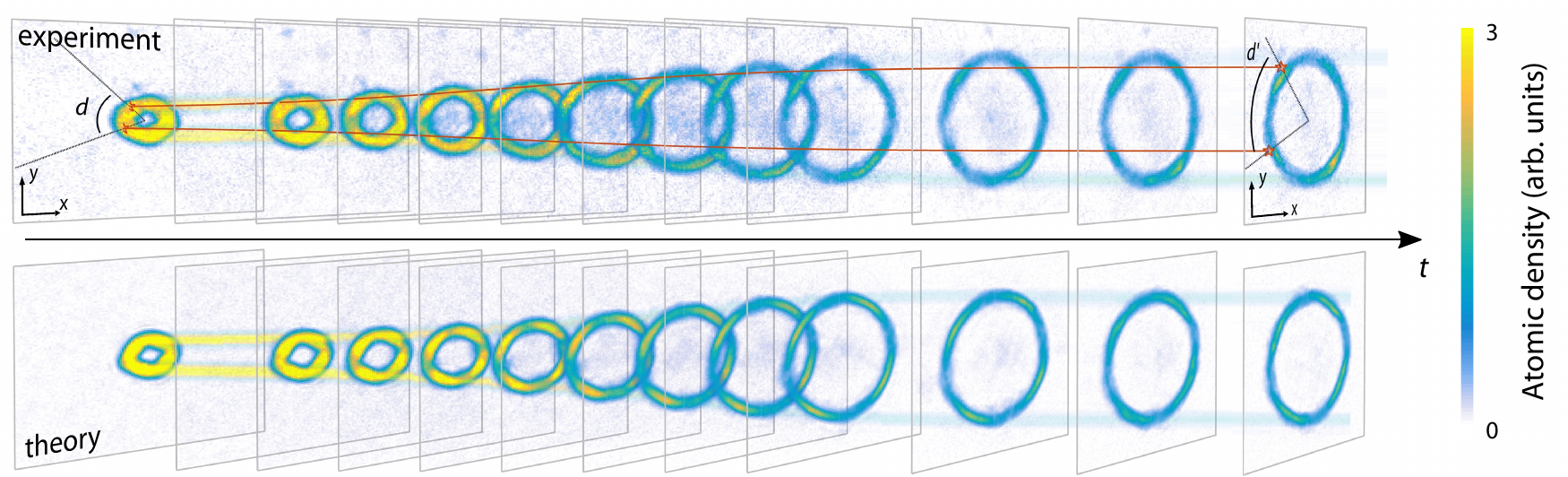}
	\caption{\label{fig:intro} Measured (top) and simulated (bottom) supersonic expansion of the ring with scale factor $a = R_f/R_i = 4.1(3)$, where $R_f=46.4(1.4)\um$ ($R_i = 11.3(4)$~$\mu$m) is the final (initial) radius~\cite{Note3}. An initial distance $d$ transforms into a larger distance $d'$.  The time elapsed in the figure is approximately 15~ms.}
\end{figure*}

\begin{figure*}
	\center
	\includegraphics{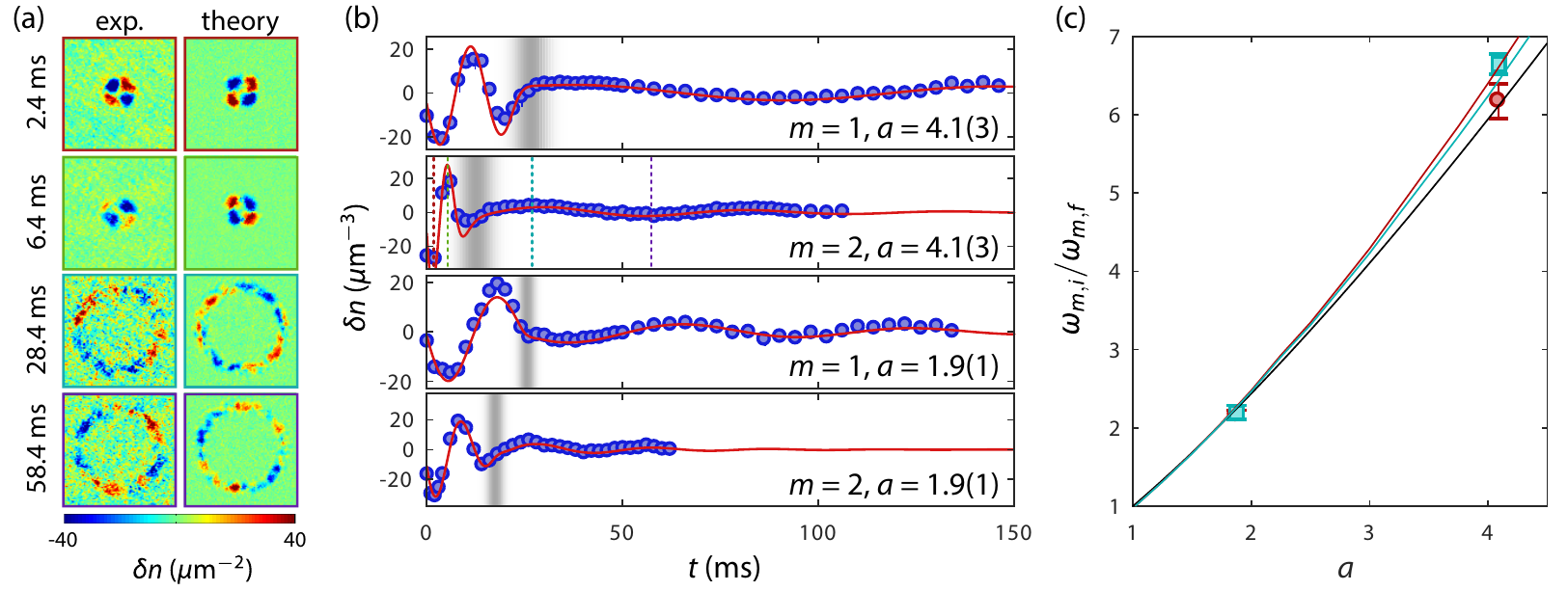}
	\caption{\label{fig:redshift} Redshift of long-wave excitations.  (a) Atomic density difference $\delta n$ at various times for both experiment and simulation for a mode number $m=2$ and scale factor $a=4.1$.  Density scale of images after expansion are multiplied by 10.  (b) Phonon amplitude vs. time for various $a$ and $m$.  The grey bands indicate the time during which the BEC is inflated; their intensity denotes the expansion velocity relative to that expansion's maximum.  Vertical dashed lines in the $m=2,a=4.1$ panel indicate the times shown in (a).  (c) Ratio of initial to final frequency vs. scale factor $a$.  Red circles indicate $m=1$ modes; cyan squares, $m=2$.  Solid, black curve is the $a^{9/7}$ expectation, and colored curves (with mode numbers matching points) are the result of full Bogoliubov calculation.}
\end{figure*}

\newpage
\twocolumngrid
{\it Phonon redshift --} To study the red shifting of phonons, we first imprint a standing wave phonon excitation on the background BEC.  During expansion, these effectively one-dimensional azimuthal phonons are redshifted, i.e., their wavelength grows as shown in Fig.~\ref{fig:redshift}a for both experiment and theory.  These images show the oscillation of a standing-wave phonon, constructed by perturbing the condensate with a potential of the form $\sin(m\theta)$, where $m$ is the integer azimuthal mode number of the phonon.  The (approximate) axisymmetry implies that $m$ is conserved, in analogy with conservation of the comoving wavevector in cosmology.  The phonon wavelength is therefore stretched by a factor $a = R_f/R_i$, the ratio of the geometrical radii of the expanding ring.  This is related to the usual redshift parameter $z$ through $a = z+1$.

Figure~\ref{fig:redshift}b shows the measured phonon amplitude $\delta n$ vs. time for various $a$ and $m$ and clearly shows a shift in the frequency.  (In this paper, we measure frequency and time in the laboratory frame, as opposed to using the comoving proper time as defined by the effective metric, Eq.~\ref{eq:metric}.)  To measure the frequency shift, we fit the oscillation before and after expansion to extract $\omega_{m,i}/\omega_{m,f}$, shown in Fig.~\ref{fig:redshift}c.  At any given time, the phonon oscillation frequency is $\omega(t) = c_\theta(t) m/R(t)$, where $c_\theta(t)$ is the azimuthal speed of sound at time $t$.  As the ring expands, both the atomic density and $c_\theta$ decrease.  For the combination of harmonic confinement in the vertical direction and roughly quartic confinement in the radial direction, we find $c_\theta\propto R^{-2/7}$.  The solid, black curve shows the resulting $\omega_{m,i}/\omega_{m,f} = a^{9/7}$ scaling; a full Bogoliubov calculation, with the azimuthally averaged potential, is shown as the solid, colored curves.

We understand the phonon's behavior during the expansion epoch in terms of a 1D equation for the phonon amplitude $\chi_m$,
\begin{eqnarray}
	\label{eq:redshift_w_damping}
	\frac{\partial^2 \chi_m}{\partial t^2} + \left[2\gamma_m(t) + \frac{\dot{R}}{R}\right]\frac{\partial \chi_m}{\partial t} + [\omega(t)]^2 \chi_m = 0\ ,
\end{eqnarray}
where $\delta n = (\hbar/U_0) \partial \chi_m/\partial t$, $U_0 = 4\pi \hbar^2 a_s/M$, $a_s$ is the $s$-wave scattering length, and $M$ is the mass of an atom.  (See Appendix~\ref{app:redshift} for the derivation.)  There are two  contributions to the damping of the amplitude.  The first damping term, $\gamma_m$, is phenomenological, but independently measured~\cite{Note4}.  The second, $\dot{R}/R$, is analogous to the ``Hubble friction'' in cosmology, which damps fields with frequencies $\omega\lesssim \dot{a}/a$.  In the present case, the Hubble friction has the largest impact when for supersonic expansion, i.e., when $\omega \lesssim \dot{R}/R$ or $m c_\theta\lesssim \dot{R}$.

For our expansions, we expect that the Hubble friction will play a role, particularly for the $a=4.1, m=1$ expansion where $\dot{R}/R\gtrsim 1.5\omega$.  (At maximum velocity, $\dot{R}/R \approx 3 \gamma_m$ for $m=2$ and  $\dot{R}/R \gtrsim 20 \gamma_m$ for $m=1$, but this occurs only during the short expansion epoch.)  The Hubble friction term changes the phase and amplitude of the phonon oscillation after expansion.  However, because the observed density difference $\delta n$ is proportional to $\partial \chi_m/\partial t$ (see Appendix~\ref{app:redshift}), the predominant difference in observed amplitude before and after expansion results from the change in $\omega$.  To search for the  Hubble friction term, we fit all the data simultaneously to Eq.~\ref{eq:redshift_w_damping}, taking $\dot{R}/R\rightarrow \gamma_H \dot{R}/R$, where $\gamma_H$ is a tunable parameter.  While the best-fit value $\gamma_H=0.55(21)$ indicates the presence of Hubble friction, the deviation from unity suggests that other effects like azimuthal asymmetry and non-zero annular thickness also affect the phonon amplitude~\cite{Note3}.  For GPE simulations of the expansion of an azimuthally symmetric, thin annulus ring with a potential of a similar functional form, Eq.~\ref{eq:redshift_w_damping} is an accurate description of the phonon evolution  (see Appendix~\ref{app:redshift}).  

\begin{figure}
	\center
	\includegraphics{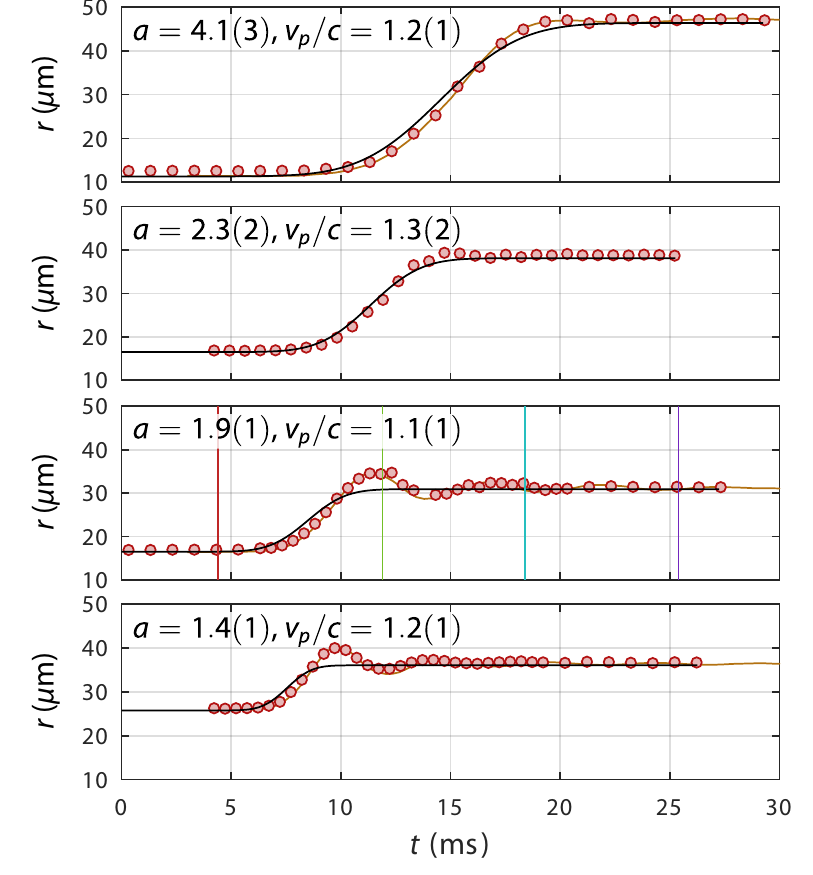}
	\caption{\label{fig:exp_profiles} Mean radius of the ring vs. time for select expansions.  Black solid curves show the radius of the potential, red circles show experimental data and orange curves show simulation results.  The vertical lines for $a=1.9(1)$ correspond to times shown in Fig.~\ref{fig:radial_exc_and_phonon}.}
\end{figure}

\begin{figure*}
	\center
	\includegraphics{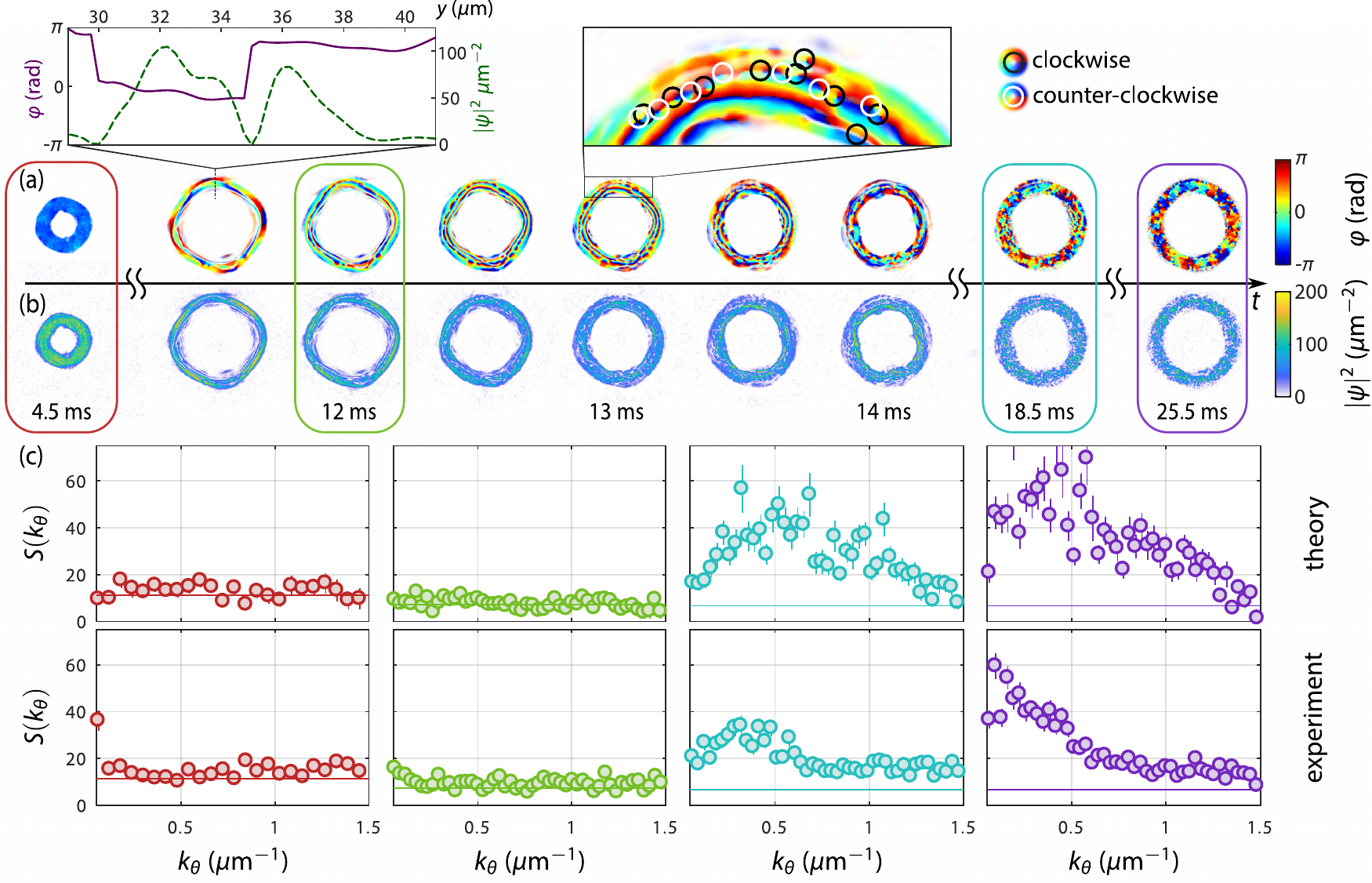}
	\caption{\label{fig:radial_exc_and_phonon} Dynamics in the radial direction and generation of azimuthal excitations for a scale factor of $a=1.9(1)$.   (a--b) A single realization of the simulated BEC wavefunction's phase (a) and magnitude $|\psi|^2$ (b) vs. time.  (Left inset) Cross section along $\hat{y}$ of the phase and density at $t=11.5$~ms.  (Right inset) Zoomed section of the phase profile at $t=13$~ms with the location of vortices highlighted. (c) Azimuthal static structure factor.  The color indicates the time, and matches the the vertical lines in Fig.~\ref{fig:exp_profiles} and the rectangles in (a) and (b).  The horizontal lines indicate the imaging detection threshold (see text).}
\end{figure*}

{\it Radial dynamics --} The preceding 1D discussion (based on Eq.~\ref{eq:redshift_w_damping}), rested on the assumption that the background BEC contained no transverse dynamics.  Perhaps the first indication of additional dynamics is visible in evolution of the ring-BEC's radius, shown by the red symbols in Fig.~\ref{fig:exp_profiles}.  As indicated by the oscillations around the trap's mean radius (black curves), the BEC is excited after the potential has reached its final value.  The amplitude of the oscillation can be estimated based on a simple harmonic oscillator model, where the oscillator is the first radial phonon mode and forces applied are due to the expansion of the confining potential.  These oscillations decay rapidly, typically within a few oscillation periods for all scale factors and expansion velocities studied.  If the trap were perfectly harmonic, this center-of-mass oscillation should be long-lived.  In reality, our trap is more flat-bottomed, is anharmonic, and is not axially symmetric.

To understand this rapid decay, we show the atomic density and phase of a simulated Bose-Einstein condensate without an imprinted phonon during the first few oscillations after expansion in Fig.~\ref{fig:radial_exc_and_phonon}a and b, respectively.  At $t=10$~ms, the condensate reaches the far end of the potential and begins to turn around.  At $t=11.5$~ms, the condensate phase is approximately flat radially, with the exception of a discontinuity of $\approx \pi$ in the center of the annulus.  This standing wave has nodes in the atomic density with corresponding $\pi$ phase jumps, effectively imprinting a dark soliton onto the BEC~\cite{pethick2002bose,Frantzeskakis2010a}.  This process is analogous to the creation of solitons upon Bragg reflection in an optical lattice~\cite{Scott2002} or reflection of a condensate off of a tunnel barrier~\cite{Martin2007}.  (Unfortunately, due to imaging limitations, we are unable to resolve solitons or other similarly-sized structures in the experiment.)

The number of solitons $N_s$ created from the decay of the  radial mode can be estimated by comparing the energy per particles contained the radial excitation to the energy per particle of a soliton ($\epsilon_s \approx 4\hbar c/3 R_T$, where $R_T$ is the annular width of the ring).  The amplitude of the radial excitation $\chi_r$, while calculable analytically, is a complicated function that depends exponentially on the adiabaticity of the expansion relative to the frequency of the radial mode $\omega_r$.  (Assuming a box-like potential in the radial direction implies $\omega_r \approx \pi c/R_T$.)  The adiabatic condition then demands, in our system, $\dot{R}$ must nearly be supersonic to produce solitons, i.e., $v_p \gtrsim 0.8c$~\cite{Note5}.

{\it Turbulence and reheating-} Dark solitons are unstable in condensates of more than one dimension.  They suffer from a ``snake instability'' causing the soliton to first undulate and then fragment into vortex dipoles~\cite{Kuznetsov1988,Kamchatnov2008,Cetoli2013}.  As shown by our numerics in Fig.~\ref{fig:radial_exc_and_phonon}a, the undulation is underway by 12.5~ms and the fragmentation into vortices is mostly complete by 14~ms.  Theoretical estimates for a single soliton in a harmonically confined BEC suggest that the snake instability will result in $N_\text{vd,1} \approx 2\pi R/8\xi$ vortex pairs, where $\xi = \sqrt{\hbar^2/2M\mu}$ is the local healing length within the bulk of the condensate and $\mu$ is the chemical potential~\cite{Toikka2013}.  For the present case, this corresponds to $N_{vd,1}\approx 50$ vortex pairs.  At $t=13$~ms in Fig.~\ref{fig:radial_exc_and_phonon}b, the single soliton has decayed into $\approx 6$ pairs over an angle $\approx 45^\circ$  near the top of the ring.  This corresponds to roughly 48 vortex pairs around the full ring.  These vortex pairs then form a highly turbulent state.

We experimentally observed the fingerprints of this process through the structure factor $S(k_\theta)$, a measure of the spatially structured density fluctuations (i.e., azimuthal phonons) with wavevector $k_\theta=m/R$.  For both experiment and theory we extracted $S(k_\theta)$ by first evaluating the one-dimensional density $n_\text{1D}(\theta)$ around the ring to obtain the density fluctuations $\delta n_{\rm 1D}(\theta) = n_{\rm 1D}(\theta) - \langle n_{1 D}(\theta) \rangle$, where $\langle \cdots \rangle$ denotes the average over many realizations.  The structure factor is  
\begin{equation}
	S(k_\theta) = \left< \left|\int \delta n_\text{1D}(\theta) e^{-i k_\theta R \theta}R \ d\theta \right|^2 \right>. 
\end{equation}
Theoretical structure factors are shown in the top row of Fig.~\ref{fig:radial_exc_and_phonon}c; experimental structure factors are shown in the bottom row Fig.~\ref{fig:radial_exc_and_phonon}c.  The colors in Fig.~\ref{fig:radial_exc_and_phonon}c identify the times at which the structure factors were evaluated.  The density obtained from experiment has limited spatial resolution, is impacted by imaging aberrations, and has additional noise from the partial transfer absorption imaging process~\cite{Ramanathan2012}.  For these reasons, we first corrected for imaging aberrations (see Appendix~\ref{sec:exp_details}) and identified the detection threshold (shown by the horizontal lines).  We used the numerical simulations (which include the same aberrations) to verify the correspondence between the corrected value of $S(k_\theta)$ based on simulated imaging to the value of $S(k_\theta)$ calculated from the simulated atomic density.  These agree for values of $S(k_\theta)$ above the detection threshold.

As shown by the $S(k_\theta)$ snapshots, the structure factor starts at our detection threshold~\cite{Note6}.  After expansion and during the soliton's initial formation ($t=12$~ms), $S(k_\theta)$ maintains this value, indicating that this state does not differ significantly between realizations.  When the soliton begins to break apart at $t=13$~ms, a small peak, still below our detection threshold, appears in the simulations near $k\approx1$~$\mu$m$^{-1}$ (not shown).  This corresponds roughly to the wavenumber of the snake instability, $k \approx 2\pi/8\xi\approx1.3$~$\mu$m$^{-1}$.  As the turbulent state develops, this peak grows and shifts to lower $k$, becoming detectable at $18.5$~ms and becoming larger at $22.5$~ms.  The shift to lower $k_\theta$ is expected because of the inverse cascade that occurs in two-dimensional turbulence~\cite{Kraichnan1967}.

\begin{figure}
	\center
	\includegraphics{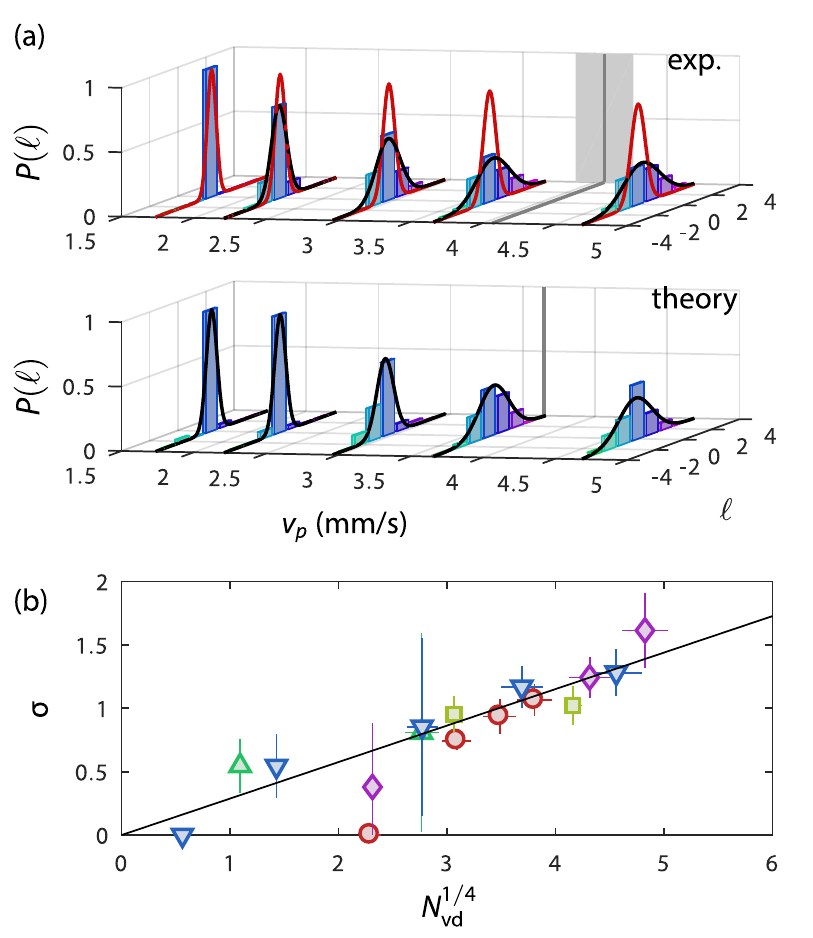}
	\caption{\label{fig:winding_number} (a) Measured winding number distributions for $a=1.4(1)$ (histograms) with Gaussian fits (black curves) for the experiment (top) and simulations (bottom).  The gray, vertical line and bar shows, for comparison, the speed of sound and its uncertainty.  The red Gaussians show the expected distribution from the horizon model (see text). (b) Width of the winding number distribution $\sigma$ vs. the number of vortex dipoles predicted by the soliton model, $N_\text{vd}$.  The red circles denote $a = 1.31(9)$ [$R_i=16.7(8)$~$\mu$m, $R_f=22(1)$~$\mu$m], yellow squares $a = 1.9(1)$ [$R_i=16.7(8)$~$\mu$m, $R_f=31(2)$~$\mu$m], green triangles, $a = 2.3(2)$ [$R_i=16.7(8)$~$\mu$m, $R_f=39(2)$~$\mu$m] and blue inverted triangles, $a=1.4(1)$ [$R_i=25(1)$~$\mu$m, $R_f=35(2)$~$\mu$m].  The purple diamonds also show $a=1.4(1)$, but with a ring that is twice as wide.  Solid line shows $\sigma = N_\text{vd}^{1/4}$.}
\end{figure}

{\it Stochastic persistent currents --} While most of the vortex dipoles recombine and produce lower energy phonons, some of the vortex dipoles manage to break apart and become free vortices.  If one of the free vortices slides into the center of the ring and one leaves the outside of the ring, then the overall phase of the ring slips by $2\pi$ and the winding number $\ell$, quantifying the persistent current state of the ring, changes by one~\cite{Anderson1966}.  Indeed, we observe stochastic persistent currents in the ring after expansion in both the experiment and simulation. Figure~\ref{fig:winding_number}a shows the resulting distributions of winding numbers for various speeds of expansion for $a = 1.4(1)$ [$R_f=35(2)\um$ and $R_i=25(1)\um$].  

Evidence for this process can be found by studying the width of the winding number distributions for expansions with different $a$ and $t_\text{exp}$.  The number of vortex dipoles produced from $N_s$ solitons would be $N_\text{vd}\approx N_s (2\pi R_f/8\xi)$.  The measured distribution widths collapse reasonably well when plotted versus $N_\text{vd}^{1/4}$, as shown in Fig.~\ref{fig:winding_number}b.  The 1/4 may result from some combination of the stochastic nature of dipole dissociation and recombination, the interaction-driven dynamics of dipoles and free vortices in a turbulent fluid, and the random phase-slip process.

One may question whether the appearance of the winding number might involve another cosmological phenomenon: the presence of sonic horizons.  If we assume the speed of sound sets a limit on the speed at which information can travel through the condensate, the rapid supersonic expansion should create regions of condensate that are causally disconnected.   The typical horizon distance established during the expansion would be given by,
\begin{equation}
	R_\text{hor} = 2\int_0^\tau c(t) dt \lesssim 2 c_0 t_\text{exp}
\end{equation}
where $c_0$ is the initial speed of sound~\cite{Note7}.  This leads to $N_R \gtrsim 2\pi R_f/R_\text{hor} \approx \pi R_f/c_0 t_\text{exp}$ disconnected regions.  If these regions' phases evolve at different rates and become sufficiently randomized, then when the regions recombine, they can form a topological excitation in the form of a persistent current~\cite{Zurek1985,Scherer2007}.  The probability for a given persistent current is then given by the geodesic rule~\cite{Bowick1994,Scherer2007,Note8}.  The red Gaussians in Fig.~\ref{fig:winding_number}a show the expected distributions resulting from this horizon model, which disagree with the experiment.  Moreover, simple estimates for the phase fluctuations present in our condensate are a factor of 25 too low to sufficiently randomize the phase during expansion.  Future studies using condensates of lower density could see this effect, as the phase fluctuations will be larger.

{\it Discussion and Outlook --} In this work, we explored the physics of a rapidly expanding Bose-Einstein condensate.  We observed the redshifting of phonons during this rapid expansion, which has clear analogs in cosmological physics.  After expansion stops, the condensate reheats through the creation and subsequent destruction of dark solitons, producing a highly turbulent state. This process leads to the creation of global topological defects (i.e., persistent currents), which at first might be thought to arise due to the presence of cosmological horizons, but actually result from the vortices produced when the solitons break apart. 

While we see evidence for Hubble friction in our system, future studies should be able to more precisely measure its influence during the expansion of the phonon modes.  In particular, by varying $\dot{R}/R$, one could more easily distinguish between the Hubble friction and other damping effects.  One could also contract the ring rather than expand it.  Because the Hubble friction is not dissipative and is reversible, such a contraction should cause amplification of the phonon mode amplitude.

The process of expansion, which presumably cools the azimuthal degrees of freedom of the condensate, followed by the increase in azimuthal excitations (Fig.~\ref{fig:radial_exc_and_phonon}c-d) as the radial mode decays, is reminiscent of the reheating process in the early universe.  At the end of inflation in the universe, the energy contained in the homogeneous mode of the quantum field that drove inflation, the inflaton, decayed into inhomogeneous excitations.  It is not known how this occurred.  In the simplest model, the inflaton oscillated around the minimum of its potential, decaying into lower energy particles~\cite{Kofman1994}, whereby the radial mode couples directly to lower energy azimuthal phonon modes.  However, the decay of the radial mode through this process is expected to be much slower ($\approx 1$~s$^{-1}$, using a calculation similar to that found in Ref.~\cite{Pitaevskii1997a}) compared to the observed decay of the radial mode through soliton and vortex excitations ($\approx 100$~s$^{-1}$).  Future studies using a ring with stronger radial confinement should suppress the non-linear excitations and enhance the direct coupling.  Other models are non-perturbative and include self-interactions in the inflaton field that can lead to turbulent cascading~\cite{Felder2001,Kofman2001,Copeland2002,Gleiser2010,Amin2012,Amin2014},  much like the turbulence we observe here.

Perhaps surprisingly, the long-wavelength azimuthal phonon mode is redshifted in simple way (Fig.~\ref{fig:redshift}), despite the complex dynamics occurring in the underlying BEC state.  This survival has a direct analogy in inflationary cosmology. During inflation, vacuum fluctuations were redshifted to large length scales and amplified. The subsequent preheating and thermalization processes took place on shorter length scales, yet the resulting thermal state was modulated by the long-wavelength amplified vacuum fluctuations.  This process gave rise to the large-scale structure we observe today in the universe.

In addition to the possibilities described above, we anticipate that with new developments, other interesting cosmological phenomena might be realized with expanding condensates.  First, with improved imaging that captures the initial (quantum and/or thermal) fluctuations, one could observe effects related to the scaling of the vacuum.  In particular, one could observe cosmological particle  production~\cite{Fedichev2003,Barcelo2003,Fedichev2004,Fischer2004,Calzetta2005,Jain2007a,Prain2010}.   Second, a ring with stronger radial confinement will suppress transverse excitations, revealing the physics arising from the recombination of causally-disconnected regions.  Given these possibilities, we believe an expanding ring BEC could provide an interesting laboratory test bed for cosmological physics.

\begin{acknowledgments}
The authors thank J. Ho for initial discussions and a careful reading of the manuscript.  We thank W.D. Phillips, E. Goldschmidt, M. Edwards and N. Proukakis for useful discussions.  We thank the anonymous referees, whose comments greatly improved the manuscript.  This work was partially supported by ONR and the NSF through the PFC at the JQI.  TJ was supported in part by NSF grants PHY-1407744 and PHY-1708139.  IBS was partially supported by the AFOSR's Quantum Matter MURI and NIST.
\end{acknowledgments}

\appendix

\section{Experimental Details}
\label{sec:exp_details}
Our experimental setup consists of a BEC of $^{23}$Na atoms in an optical dipole trap (ODT).  Our BECs are created using standard laser cooling techniques, followed by evaporation in first magnetic then optical dipole traps.  In this experiment, we work with BECs with between $1\times10^5$ and $4\times10^5$ atoms.  For measurement, we use partial transfer absorption imaging (PTAI)~\cite{Ramanathan2012}.

The final stage of evaporation begins when thermal atoms are loaded into a combination of the vertical trap and dimple trap.  Vertical confinement is created using a blue-detuned ($532\nm$), TEM$_{01}$ beam, tightly focused to create two parallel sheets of light with a dark region in between.  The dimple trap is a red-detuned ($1064\nm$) Gaussian beam with $1/e^2$ diameter $\approx 50$~$\mu$m that provides the initial confinement in the horizontal plane.  Forced dipole evaporation occurs by lowering the intensity of both the blue-detuned vertical confinement beam and the red-detuned Gaussian beam until the condensate reaches a condensate fraction $>95$~\%.  We estimate the initial temperature to be of the order of 50~nK by extrapolation of the evaporation process~\cite{Kumar2017}.   The final vertical trapping frequency is 650(4)~Hz.  The atoms are then adiabatically transferred to the initial trap for the experiment.

To create the initial ring (or target) trap, we use a direct intensity masking technique to create the blue-detuned ($532\nm$) trap in any shape.  Details of this technique can be found in Refs.~\cite{Lee2014,Kumar2016}.  Briefly, this approach images the face of a digital micromirror device (DMD) that is illuminated by a blue-detuned Gaussian beam and imaged onto the atoms.  The pattern written onto the DMD is then transferred onto the potential experienced by the atoms.  Using this technique, we can form fully-dynamic potentials in the shape of rings (with radii between 10~$\mu$m and 45~$\mu$m) and target shaped traps (for measuring the persistent current state of the ring).     The $1/e^2$ radius of the Gaussian beam that illuminates the DMD is 130(10)~$\mu$m in the plane of the atoms. 

Nominally, the pattern written on the DMD is given by
\begin{equation}
	V_\text{DMD}(r) = \Theta([R(t)-R_T/2]-\rho) + \Theta(\rho-[R(t)+R_T/2]), \label{eq:pot_without_pert}
\end{equation}
where $\Theta$ is the Heaviside step function, $R_T$ is the ring's width, and $\rho$ is the radial coordinate.  For rings thinner than $R_T<10\um$, we apply corrections by changing $R_T$ with angle $\theta$ to make the measured $n_{1D}(\theta)$ density of the condensate more uniform.

To expand the ring, we apply a time-dependent potential using our DMD.  To minimize spurious effects related to jerk, we used a smoothly varying function of the form
\begin{equation}
	\label{eq:ring_radius}
	R(t) = \left\{ \begin{array}{lr} 
	R_i & t\leq 0 \\
	R_i + \frac{1}{2}(R_f-R_i)(1+\text{erf}\left\{\frac{1}{\beta}\left(\frac{t}{t_\text{exp}}-\frac{1}{2}\right)\right\} & \ \ \ 0<t\leq t_\text{exp} \\
	R_f & t>t_\text{exp}
	\end{array}\right. ,
\end{equation}
where erf is the error function and $\beta$ is a parameter that minimizes the jerk at $t=0$ and $t=t_\text{exp}$.  For the data reported in this paper, $\beta=0.175$, which implies that at $t=0$ and $t_\text{exp}$ the radius suddenly jumps by $\approx3\times10^{-5}(R_f-R_i)$.   The DMD is pre-programmed with individual frames with ring radii calculated using Eq.~\ref{eq:ring_radius}.  We use approximately $30$ frames spaced $\approx 300$~$\mu$s apart to encode the expansion of the ring.  Given our typical chemical potentials of $\approx 1$~kHz, this update rate is faster than all other timescales in the system.   Moreover, we checked that our results are independent of the number of frames used.  During the expansion, we increase the intensity of the trapping light to maintain constant intensity locally near the ring (compensating for the Gaussian profile of the beam illuminating the DMD).  We tune the increase in the trapping light to keep the frequency of the first radial Bogoliubov mode constant with radius.

To imprint a phonon of mode number $m$, we instantaneously change this pattern to
\begin{eqnarray}
	V_\text{DMD}(r,\theta) & = & \Theta\left(\left[R(t)-\frac{R_t}{2}\right]-\rho\right) + \Theta\left(\rho-\left[R(t)+\frac{R_t}{2}\right]\right) \nonumber \\
	& & + \frac{\lambda}{2} \Theta\left(\rho-\left[R(t)-\frac{R_t}{2}\right]\right)\Theta\left(\left[R(t)+\frac{R_t}{2}\right]-\rho\right) \nonumber \\
	& & ~~~~~~ \times \left[1+\sin(m\theta)\right]. \label{eq:pot_with_pert}
\end{eqnarray}
Here $\lambda = 0.6$ is a parameter that describes the size of the perturbation relative to overall potential depth.  One cannot generate the necessary values between 0 and 1 to produce the potential described by Eq.~\ref{eq:pot_with_pert} with a binary DMD device.  To get the necessary grayscale to create the potential,  the DMD is demagnified in order to make its pixel size ($\approx 0.5\um$ in the plane of the atoms) be much smaller than the aberrated point spread function ($\approx 4\um$ $1/e^2$ full-width) of our imaging system.  We then use halftoning to create the necessary grayscale effect.  Ref.~\cite{Kumar2016} contains more details about this imprinting process.

To measure the normalized phonon amplitude after imprinting, we first measure the 2D density {\it in situ} with ($n_\text{2D}(\rho,\theta)$) and without ($n_\text{2D,0}(\rho,\theta)$) the phonon imprinted.  We then integrate over the radial dimension to obtain the 1D density around the ring, e.g., $n_\text{1D}(\theta) = \int n_\text{2D}(\rho,\theta)d\rho$.  To obtain the normalized 1D density, we compute $n_\text{1D}(\theta)/n_\text{1D,0}(\theta)$.  The data are fit to $a_m \sin(m(\theta+\theta_0)$ at each time to extract the normalized amplitude of the phonon $a_m(t)$.  The offset angle $\theta_0$ is set by the imprinting process.  Finally, we turn $a_m(t)$ into real phonon amplitude $\delta n(t)$ by multiplying by the total number of atoms and dividing by the estimated Thomas-Fermi volume of the condensate $V_{TF}$.  Here, we have made two implicit assumptions.  First, we have assumed that the phonon's amplitude is independent of $\rho$ and $z$, which is valid when the thickness of the annulus is small compared to its radius.  (See Appendix~\ref{app:redshift} for details.) Second, we have assumed that the Thomas-Fermi volume scales in the experiment according to how it would in a potential that is quartic in $\rho$ and harmonic in $z$.  With the same assumptions on the potential, the predicted frequency shift scales as $a^{9/7}$, which agrees rather well with the experiment (Fig.~\ref{fig:redshift}c).  We also note that an incorrect estimate of the original Thomas-Fermi volume would lead to a common scaling of the phonon amplitude at all later times (before and after expansion), which would not lead to any change in either the fitted frequency shift or Hubble friction. 

Calibration of the aberrations in our imaging system is necessary in order to accurately measure the correlation function $S(k)$.  Conveniently, PTAI allows us to accurately calibrate our imaging system's sensitivity to density structures with wavevector $k$.  When the transfer fraction $f$ is low ($f\ll 1$), quantum shot noise is added and dwarfs the thermal and quantum fluctuations inherent to the condensate.  This additional noise is white over all $k$, thus allowing for accurate calibration.  To calibrate, we measure $n_{1D}(\theta)$ as described above and then construct $S(k)$ as described in the main text.  To compensate for the our imaging system's degraded performance at larger $k_\theta$, we minimize the functional $(\frac{S(k)}{C(k)} - \frac{1}{f})^2$ using the tunable parameters $k_1$, $p_1$, $k_2$ and $k_3$ contained within the correction function:
\begin{equation}
	C(k) = \frac{1}{\sqrt{1 + |k/k_1|^{p_1}}}  \frac{1}{\sqrt{1 + |k/k_2|^{p_2}}}.
\end{equation}
The experimentally determined parameters are $k_1 = 0.34(2)$~$\mu$m$^{-1}$, $p_1 =3.4(2)$, $k_2=1.50(4)$~$\mu$m$^{-1}$ and $p_2 = 15(6)$.

To measure the persistent current state, we form a trap with a ring and a concentric, central disc (i.e., a target symbol) and use the interference between the two in time-of-flight to determine the winding number~\cite{Eckel2014,Corman2014}.   To produce acceptable interference fringes for readout, the disc must also be expanded.  This is done adiabatically over 25~ms with 40~frames.  Expansion of the ring produces a host of excitations, including phonons, vortices in the annulus, and persistent currents.  To accurately measure the persistent current with the least amount of interference from other excitations, we let the ring equilibrate for about 5~s.  During this period, the intensity of light is ramped to $\approx60$~\% of its value at the end of expansion to force evaporation of high energy excitations.

\section{Stochastic-Projected Gross-Pitaevskii Calculations}
\label{app:spgpe}

To explore the behavior of our system numerically, we conducted simulations of the stochastic projected Gross-Pitaevskii equation~\cite{Rooney2012,Bradley2015}.   This numerical framework extends the ordinary Gross-Pitaevskii equation to non-zero temperature, adding on fluctuations to the BEC ground state.  While described in detail in the aforementioned references, we will briefly describe the technique here.  In this formalism, the wavefunction of the BEC with fluctuations evolves in a ``coherent'' region -- defined as the region of Hilbert space spanned by the state vectors that impact the dynamics of the BEC coherently.  The BEC wavefunction in this C-region evolves as
\begin{equation}
	(S) d\psi = d\psi_H+d\psi_G+(S)d\psi_M
\end{equation}
where $(S)$ denotes Stratonovich integration and
\begin{eqnarray}
	d\psi_H & = & \mathcal{P}\left\{-\frac{i}{\hbar}\mathcal{L}\psi dt \right\} \label{eq:ham_term} \\
	d\psi_G & = & \mathcal{P}\left\{\frac{G(\mathbf{r})}{k_B T}(\mu-\mathcal{L})\psi dt + dW_G(\mathbf{r},t)\right\} \label{eq:growth_term}
\end{eqnarray}
Here, $\mathcal{L} = H_{sp} + U_0 |\psi|^2$ is the driver of Hamiltonian evolution and $H_{sp} = p^2/2M+V$ is the single particle Hamiltonian.  The equation for $d\psi_G$ represents growth of population in the C-region from particles colliding in the incoherent (I) region.  Here, $G(\mathbf{r})$ is a coefficient that sets the strength of both terms in Eq.~\ref{eq:growth_term}, where the first term is the damping term and the second is the growth term where $dW_G$ describes a random noise seeded according to $\langle dW_G^*(\mathbf{r'},t')dW_G(\mathbf{r},t)\rangle = 2 G(\mathbf{r})\delta(\mathbf{r}'-\mathbf{r})dt$.  For this work, we neglect terms where there is an exchange of energy and momentum between the C and I region without exchange of particles~\cite{Rooney2012}.  Finally, the projector operator $\mathcal{P}$ continually projects the wavefunction into the C region.

From an implementation perspective, this involves taking a Gross-Pitaevksii equation solver and adding a noise term, and appropriately calculating the damping factor $G(\mathbf{r})$, which is assumed to be constant.  Our calculations are done in a Cartesian coordinate system.  We apply the projection operator in momentum space, with a cutoff $k_c\approx \pi/\delta x$, where $\delta x$ is the spacing between points in the grid.   

To accurately capture the potential, we simulate the imaging process that is used to make the potential.  We reproduce the image that is patterned on the DMD and simulate imaging using Fourier imaging techniques. The aperture function of the imaging system that relays the image from the DMD to the atoms is crucial in order to accurately replicate the potential at the atoms.  In the experiment, the same imaging system that is used for making potentials is also used for imaging of the atoms.  By measuring density-density correlations in a simple-connected thermal gas with noise dominated by quantum shot noise (by using $f\ll1$), we can extract the even (symmetric under parity reversal) aberrations~\cite{Hung2011a}.  To extract the odd aberrations, we use a less precise means.  A second DMD in the Fourier plane of the imaging system can be used to measure the geometric spot diagram, yielding another, independent means of obtaining the aperture function.  The two methods are in agreement.  We use the even aberrations from the correlations and the odd aberrations for the spot diagram technique to construct the aperture function. Finally, we neglect variations in the intensity of the trapping light caused by unwanted scattering along the optical path (i.e., speckle).  Because the atoms seek the darkest part of the imaged potential (the trap is blue detuned), the effect is minimized.   Speckle from the surface of the DMD is eliminated by imaging.

We combine the aperture function with the Gaussian beam.  We assume the beam is perfectly Gaussian and is centered on the DMD.   Because the beam portion of the potential tends toward zero as $r\rightarrow\infty$, we establish a low energy potential floor at large radius.  This cutoff is determined by the minimum value of the imaged and aberrated potential between $R + \frac{3}{2}R_t<\rho <R + 1.1\times \frac{3}{2}R_t$.  This prevents spurious effects like the appearance of additional BEC components out at large radius. 

The resulting potential is complicated and not easily expressible in an analytic form.  However, when azimuthally averaged, the potential has the form $V = \frac{1}{2}M\omega_r^2(\rho-R(t))^2 + \lambda\tilde(\rho-R(t))^4$, with $\omega_r \approx 2\pi\times 100$~Hz and $\lambda/h \approx 0.8$~Hz~$\mu$m$^{-4}$.  Because most ($\approx 90$~\%) of the confinement comes from the quartic term, it is generally acceptable to neglect the quadratic term for the purposes of calculating static properties like the initial and final $\mu$ and $c$.

Given that some atoms are lost during the expansion, we must also include absorbing boundary conditions in the simulation.  We do this by including a potential
\begin{equation}
	V_a = \left\{\begin{array}{cr} 0 & \rho<R_c \\
								-i V_a e^{-w_a/(\rho-R_c)} & \rho>R_c
								\end{array} \right.,
\end{equation}
where $R_c$ is a radial cutoff at which the potential turns on, $V_a$ is the amplitude of the potential, and $w_a$ is a parameter that controls the width.  The function is a non-analytic, continuously differentiable function that minimizes the reflections from the absorbing boundary.  We chose $w_a\approx 25$~$\mu$m and $V_a/h\approx 1$~kHz.  This generally results in the best absorption and the least reflection.  

With all of these components, the simulations proceed as follows.  We first find the equilibrium state by evolving the SPGPE (without $V_a$) for approximately 50~ms to 100~ms using a growth and decay term that are 100 times that of the value specified by the temperature (this allows for faster equilibration times).  Second, we expand the ring according to that the same profile as seen in the experiment.  Approximately halfway through the experiment, we turn on $V_a$ to ensure that the decay of the atom number is appropriately captured.  After evolving for a total of approximately 35~ms (20~ms additional after the end of the expansion), we turn off the stochastic growth term in the SPGPE and turn on significant damping to determine whether or not a winding number is present in the condensate.  We do this approximately 25 independent times to gather statistics.

We then use the same data analysis tools used on the experimental data to extract the winding number distributions, structure factor as a function of time, and radius of the ring as a function of time.  The structure factor, as was done in the experiment, is measured relative to the mean density around the ring.  As a result, the structure factor is determined solely by the differences in density between a given simulation and the mean of all the simulations.

\section{Evolution of azimuthal phonons}
\label{app:redshift}

In this appendix, we derive the wave equation satisfied by the phonon field, and explain the origin of the redshifting and Hubble friction.  We start by noting that the Gross-Pitaevskii equation, written in terms of the density $n$ and phase $\phi$ defined through $\psi = \sqrt{n(\mathbf{r},t)} e^{i\phi(\mathbf{r},t)}$, can be expressed as an ideal fluid with an equation for continuity,
\begin{equation}
	-\frac{\partial n}{\partial t} = \frac{\hbar}{m}\nabla\cdot(n\nabla\phi),
\end{equation}
and an equation analogous to the Euler equation,
\begin{equation}
	\label{eq:euler}
	-\hbar \frac{\partial \phi}{\partial t} =  -\frac{\hbar^2}{2m \sqrt{n}}\nabla^2 \sqrt{n} + \frac{\hbar^2}{2 m }(\nabla\phi)^2 + V + U_0 n, 
\end{equation}
where $V$ is the potential for the atoms, $U_0 = 4\pi \hbar^2 a_s/M$ is the interaction constant, $M$ is the mass of an atom, and $a_s$ is the $s$-wave scattering length.  In this treatment, we neglect the quantum pressure term (the first term on the right hand side of Eq.~\ref{eq:euler}).  By linearizing the equations about the background solution $n_0$ and $\phi_0$, i.e., $n  = n_0 + n_1$ and $\phi = \phi_0 + \phi_1$, one obtains the coupled differential equations
\begin{eqnarray}
		 \frac{\partial n_1}{\partial t} & = & -\frac{\hbar}{m} \nabla\cdot\left[n_0\nabla\phi_1 + n_1\nabla \phi_0\right]  \label{eq:den_ev} \\
		-\hbar \frac{\partial \phi_1}{\partial t} & = & \frac{\hbar^2}{m}\nabla\phi_0\cdot\nabla\phi_1 + U_0 n_1. \label{eq:phi_ev}
\end{eqnarray}
(In the main text, we use $\delta n = n_1$.)  Solving the second equation for $n_1$ and inserting into the first yields a wave equation for $\phi_1$, which, expressed in covariant form, is
\begin{equation}
	\label{eq:wave_eq}
	\frac{1}{\sqrt{-g}} \partial_\mu (\sqrt{-g}g^{\mu\nu} \partial_\nu \phi_1) = 0,
\end{equation}
where $g^{\mu\nu}$ is the inverse metric, $g$ is the metric's determinant,  and $\phi_1$ is the phonon's velocity potential field.  The metric, in its most general form, is given by the line element
\begin{eqnarray}
	ds^2 & = & g_{\mu\nu}dx^\mu dx^\nu \nonumber\\
	\label{eq:general_metric}
	 & = & {c_0}\left[-c_0^2\, dt^2 +(\mathbf{dx} -\mathbf{v}_0\,dt)  
     \cdot(\mathbf{dx} -\mathbf{v}_0\,dt)\right]. 
\end{eqnarray}
Here, 
$\mathbf{v}_0=(\hbar/m)\nabla \phi_0$ is the velocity field of the background condensate, and $c_0 =\sqrt{U_0 n_0/M}$ is the speed of sound. 

In the expanding ring experiment, the potential is (approximately) axisymmetric, and is translated radially as a function of time.
In terms of cylindrical coordinates $(\rho,\theta,z)$,
the velocity
$\mathbf{v}_0$ thus has only ${\rho}$ and ${z}$ components,
with its dominant component being radial. The central radius of the ring is given by a function $\rho=R(t)$ so, assuming the velocity is purely radial, 
the line element [Eq.~\ref{eq:general_metric}] takes the form
\begin{equation}
	\label{eq:metric}
ds^2 = c_0[-c_0^2\, dt^2  +(R+\tilde\rho)^2 d\theta^2 + d\tilde\rho^2 + dz^2]
\end{equation}
where $\tilde\rho=\rho-R(t)$ is a co-moving radial coordinate.
The tensor density that directly enters the wave equation [Eq.~\ref{eq:wave_eq}] is then,
in $(t,\theta,\tilde\rho,z)$ coordinates,
\begin{equation}
	\label{eq:f}
f^{\mu\nu}:=\sqrt{-g}g^{\mu\nu}
=(R+\tilde\rho)\,{\rm diag}[-1,c_0^2/(R+\tilde\rho)^{2},c_0^2,c_0^2].
\end{equation}
The speed $c_0$ is determined by the background condensate density,
which to a first approximation follows the instantaneous ground state Thomas-Fermi distribution at all times during the expansion,
\begin{equation}
	\label{eq:thomas_fermi_sol}
	n_0 = \frac{\mu - V(\tilde\rho,z)}{U_0}
\end{equation}
This density extends out to the 
contour in the $\tilde\rho$-$z$ plane where the numerator vanishes. The chemical potential $\mu$ drops as the ring expands, so that the total number of atoms remains constant.  

In the experiment, we first apply a perturbation to a stationary condensate to excite an eigenmode of the wave equation.  An eigenmode analysis based on the methods of Ref.~\cite{Zaremba1998} will be detailed in a forthcoming paper; the essential details are presented here.  Assuming azimuthal symmetry, the eigenmodes for a thin ring have the form $\phi_1 = \chi_{klm}\eta_{klm}(\tilde\rho,z;R) e^{i(\omega t - m\theta)}$, where $\eta_{klm}(\tilde\rho,z; R)$ is a function that describes the radial ($k$) and vertical excitations ($l$) of the Bogoliubov mode when the ring has radius $R$, and $\chi_{klm}$ is its amplitude.  We denote the corresponding eigenfrequencies as $\omega_{klm}$. 

While the system may begin with only a $k=l=0$ eigenmode excited, the expansion of the ring can produce transitions into other modes. The solution at all times takes the general form $\phi_1(t,\tilde\rho,\theta,z) = \sum_{klm}\chi_{klm}(t)\eta_{klm}(\tilde\rho,z) e^{-i m\theta}$,
with all $\chi_{klm}(t=0)=0$ except for $k,l=0$ and our excited mode of interest $m$. 
Azimuthal symmetry precludes coupling between modes with different values of $m$.  Furthermore, in the thin ring limit, the coupling between different $k$ modes 
tends towards zero.  
We therefore focus here exclusively on modes that are  excited only in the azimuthal direction.  (The radial excitation which occurs when the ring expansion stops and is not relevant to the redshift is discussed in the main text.)

When $m \ll \omega_{100}/(c_\theta/R)$ and for a thin ring, $\eta_{00m}(\tilde\rho,z; R)$ is constant.   (Henceforth, we will drop the $k$ and $l$ subscripts when they are both equal to zero.)  
In this limit, the equation for modes with $k,l=0$ involves just $t$ and $\theta$ derivatives,
We can thus reduce the wave equation for these azimuthal phonon modes to a  1+1 dimensional wave equation, with an effective 
sound speed $c_\theta$. 
As in Ref.~\cite{Zaremba1998}, 
$c_\theta^2$ is given by an average over the cross section of the ring. For a thin ring this takes the form
\begin{equation}
	c_\theta^2 = \frac{1}{AM}\int\left[\mu - V(\tilde\rho,z)\right]\  d\tilde\rho dz,
\end{equation}
where the integral is over the cross section of the Thomas-Fermi wavefunction of area $A$.  For $V = \frac{1}{2}M\omega_z^2z^2 + \lambda\tilde\rho^4$ this yields
\begin{equation}
	\label{eq:eigenvalue}
	c_\theta^2 = \frac47 c^2,
\end{equation}
where $c^2 = \mu/M$ is the peak local sound speed.   By normalizing the Thomas-Fermi solution to the number of atoms $N$ one finds that $\mu\propto R^{-4/7}$, and therefore $c_\theta \propto c \propto R^{-2/7}$. 

The wave equation satisfied by 
our modes of interest, i.e., $\phi_1 = \chi_m(t) e^{i m\theta}$,  
is determined by the 
effective inverse metric density
obtained from Eq.~\ref{eq:f} by dropping the $\tilde\rho$ and $z$ components and replacing $c$ by $c_\theta$. In the thin ring limit this gives
\begin{equation}
	\label{eq:f2}
	{f}_2^{ab} = 
	{\rm diag}[-R,c_\theta^2/R]
    \end{equation}
The resulting mode equation is
\begin{equation}
	\label{eq:redshift_wo_damping}
	\ddot\chi_m + \frac{\dot{R}}{R}\,\dot\chi_m + \omega_m^2\,\chi_m=0,
\end{equation}
where $\omega_m:=mc_\theta/R$.
This is the equation of a damped harmonic oscillator, with time-dependent frequency and damping rate. We note that this particular equation does not result from the wave equation for any 1+1 dimensional metric, since there exists no
metric $g_{2ab}$ for which 
${f}_2^{ab}=\sqrt{-g}g^{ab}$. The reason is that  
the determinant of Eq.~\ref{eq:f2} is $-c_\theta^2$, whereas the determinant of 
$\sqrt{-g}g^{ab}$ is equal to $-1$ for any two-dimensional metric. 

As the ring expands, the azimuthal wavenumber $m$ is conserved, so the physical wavelength redshifts as $R^{-1}$, in analogy with the
cosmological redshift. Unlike in cosmology, the sound speed is also changing, so the frequency $\omega_m$ redshifts as $R^{-9/7}$.
In the cosmological setting, the damping term
in Eq.~\ref{eq:redshift_wo_damping}
is called the  ``Hubble friction'' term, and would be multiplied
by $3$ in three spatial dimensions.  
The Hubble damping is not actually dissipative; in fact, Eq.~\ref{eq:redshift_wo_damping}
can be obtained from the Lagrangian $L=\tfrac{1}{2} R \dot\chi_m^2 -\tfrac{1}{2}(m^2c_\theta^2/R)\chi_m^2$, which has the adiabatic invariant $R\omega_m \chi_m^2$.
To obtain Eq.~\ref{eq:redshift_w_damping} in the text, we add the phenomenological damping $\gamma_m$ observed in the experiment.

In the experiment, we measure the 
density variation $n_1$, not the phonon velocity potential $\phi_1$. The relation between
these quantities is given by 
Eq.~\ref{eq:phi_ev}.
Since $\nabla\phi_1$ is azimuthal and $\nabla\phi_0$ is radial, $\nabla\phi_0\cdot\nabla\phi_1=0$, so we have
\begin{equation}
	n_1 = -\frac{\hbar}{U_0}\dot\phi_1 = -\frac{\hbar}{U_0}\dot\chi_m e^{im\theta}\ .
\end{equation}
Hence, in the experiment, we measure the time derivative of the phonon amplitude.  

\begin{figure}
	\includegraphics{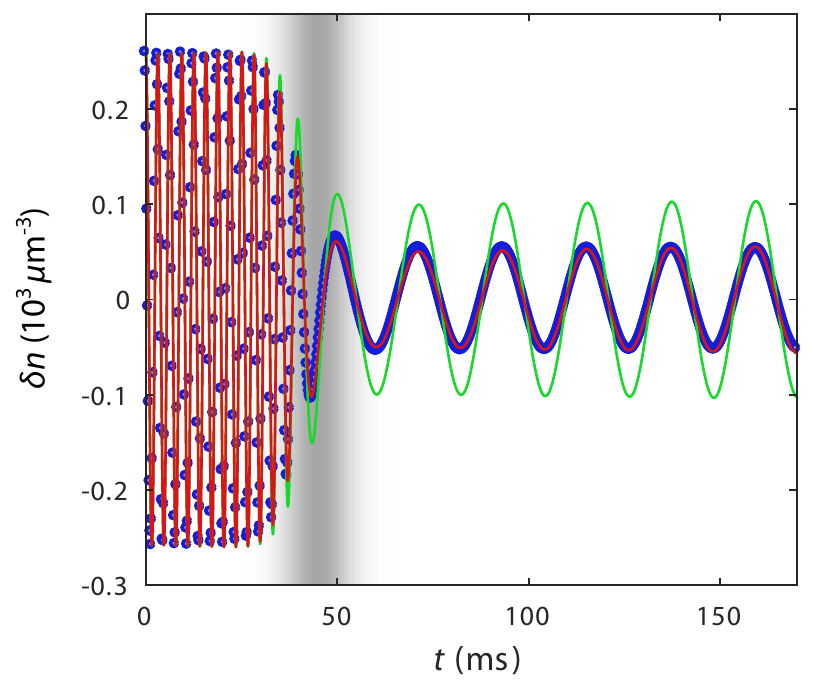}
    \caption{\label{fig:gpe_hubble} Phonon evolution in the thin ring limit.  Blue circles show 2D GPE simulation data, red (green) curve is the solution to Eq.~\ref{eq:redshift_wo_damping} with (without) the $\dot{R}/R$ ``Hubble friction" term included.  The gray band indicates the time during which the BEC is inflated; its intensity denotes the expansion velocity relative to the maximum.}
\end{figure}

We can verify that a phonon excitation does indeed obey Eq.\ref{eq:redshift_wo_damping} in a thin ring, by simulating a BEC in this regime.  Figure~\ref{fig:gpe_hubble} shows such a 2D simulation of BEC in a radially quartic potential, expanding from 10 to 40~$\mu$m in $\approx 15$~ms with $2\times10^5$ atoms.  There is no damping in this simulation; therefore, the $\gamma_m(t)$ in Eq.~\ref{eq:redshift_w_damping} is identically zero.  We choose the strength of the potential to make the initial Thomas-Fermi width be $2\um$.  As can be seen from the figure, Eq.~\ref{eq:redshift_wo_damping} accurately reproduces the behavior of the redshifted phonon, but only when the Hubble friction term is included.  Unlike the experiment, the adiabatic limit is satisfied ($\partial\omega_m/\partial t \ll \omega_m^2$) and the final amplitude is accurately predicted using the adiabatic invariant $R\omega_m \chi_m^2$. While this simulation shows the thin ring limit, we generally find that as we relax this constraint and increase the width of the annulus, the best-fit Hubble friction becomes less than unity, as might be expected from the experimental result.

\begin{figure}
	\center
    \includegraphics{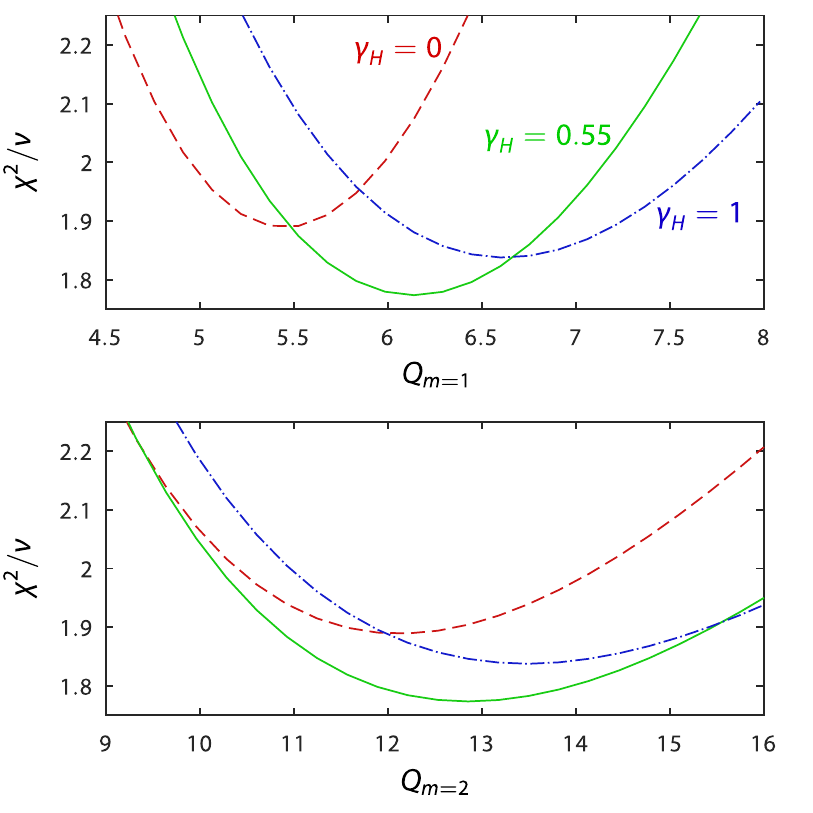}
    \caption{\label{fig:chi_sqrd_analysis} Goodness of fit $\chi^2/\nu$, where $\nu$ is the number of degrees of freedom, versus quality factors for the $m=1,2$ modes for various values of $\gamma_H$.}
\end{figure}

For the experiment, we attempt to tease out the Hubble friction by fitting it along with the other parameters of Eq.~\ref{eq:redshift_w_damping}.  These parameters are the initial amplitudes, frequencies, and phases for each of the four expansions, the quality factor for the two $m$ modes, $Q_{m} = \omega_m/2\gamma_m$, the scaling of the frequency with radius [expected to be $9/7\approx 1.2875$, the best fit value is 1.19(2)].  This fit therefore contains 15 parameters and 160 degrees of freedom.  Fig.~\ref{fig:chi_sqrd_analysis} shows the results of the reduced-$\chi^2$ fit; it shows the value of $\chi^2$ vs. both $Q_{m=1}$ and $Q_{m=2}$ in the vicinity of their best fit values for three values of $\gamma_H$, including the best fit value.  There are several interesting features.  First, the reduced-$\chi^2>1$, most likely because we do not have a good estimate of the statistical uncertainties (each point represents only four realizations of the experiment) and our model does not properly account for all the relevant effects (for example, the azimuthal asymmetry and non-zero annular thickness may play a non-negligible role in determining the phonon dynamics).  Second, $\gamma_H=1$ produces a better fit than $\gamma_H=0$, but both are improved slightly by taking $\gamma_H = 0.55$.  Third, the smallness of the change in the minimum of $\chi^2$ with $\gamma_H$ indicates our uncertainty it $\gamma_H$.  (Part of this insensitivity comes our choice, at the time of the experiment, to have $ n_1 \propto \dot{\phi} \approx 0$ during the fastest part of the expansion, thereby inadvertently minimizing the effect of the Hubble friction~\cite{Note9}.)  Taken together, the evidence is consistent with $\gamma_H=1$ but is not conclusive.

\footnotetext[1]{That is, neglecting the quantum pressure term.}
\footnotetext[2]{With perfect imaging, our potential would be flat-bottomed, but a combination of finite resolution and aberrations causes our potential to have both curvature and broken axial symmetry.}
\footnotetext[3]{Unless stated otherwise, all uncertainties and errorbars in this paper are the uncorrelated combination of 1$\sigma$ statistical and systematic uncertainties}
\footnotetext[4]{We measure $Q=\gamma_m/\omega_m\approx 6$ for all $m>1$ and $Q\gtrsim 20$ for $m=1$, indicating that the intrinsic damping $\gamma_m$ is most likely Landau damping~\cite{Pitaevskii1997a}.  $Q$ decreases slightly for increasing $m$ and $R$.  For fitting a given $m$, we take $Q$ to be constant with time.}
\footnotetext[5]{For the expansion described by Eq.~\ref{eq:ring_radius}, $\chi_r = (R_f-R_i)\exp[-(\beta \omega_r t_\text{exp})^2/4]$.  Thus, we must compare the energy of the soliton to that of the energy in the radial mode, $\frac{1}{2} m \omega_r^2\chi_r^2$.  The full adiabatic condition then becomes $v_p \gtrsim c (R_f-R_i)/R_T \times \sqrt{\pi} \left[\log\left(3\pi^2 m c (R_f-R_i)^2/\hbar R_T \right)\right]^{-1/2}$.}
\footnotetext[6]{In the simulations, $S(k_\theta)$ as computed from the density (as opposed to simulated, aberrated images) is $\approx 1$ at these early times.  This corresponds to the equilibrium value for a condensate where $k_B T \approx \mu$, $T$ is the temperature, and $k_B$ is Boltzmann's constant.}
\footnotetext[7]{The approximation of constant speed of sound is good to within 10\% because of the weak dependence of $c$ on $R$.}
\footnotetext[8]{We find through Monte Carlo simulations that the probabilities are well described by normal distributions with $\sigma\approx 0.3\sqrt{N_R}$.}
\footnotetext[9]{Our experiments were performed before we realized the significance of the Hubble friction effect, which was only later pointed out by our more enlightened theory co-author.}

\bibliography{expanding_universe}

\end{document}